\documentclass[12pt]{article}
\usepackage{a4wide}
\usepackage[utf8]{inputenc}
\usepackage[T1]{fontenc}

\usepackage{natbib}
\usepackage{amsmath,amssymb}
\usepackage[vlined,ruled]{algorithm2e}\DontPrintSemicolon
\usepackage{url}
\usepackage{comment}
\usepackage{tikz}\usetikzlibrary{arrows}
\usepackage{caption,subcaption}
\usepackage{graphicx}
\definecolor{ETHBlue}{RGB}{33,92,175}   
\definecolor{ETHGreen}{RGB}{98,115,19}          
\definecolor{ETHPurple}{RGB}{163,7,116} 
\definecolor{ETHGray}{RGB}{111,111,111} 
\definecolor{ETHRed}{RGB}{183,53,45}    
\definecolor{ETHPetrol}{RGB}{0,120,148} 
\definecolor{ETHBronze}{RGB}{142,103,19}        
\colorlet{red}{ETHRed}
\colorlet{green}{ETHGreen}
\colorlet{blue}{ETHBlue}
\colorlet{yellow}{ETHBronze}

\newcommand{\Elo}{\'El{\H o}}
\newcommand{\eighthfinal}{\mathit{L16}}
\newcommand{\quarterfinal}{\mathit{QF}}
\newcommand{\semifinal}{\mathit{SF}}

\def\cellwidth{1.9}
\def\cellheight{0.5}
\def\cellcolorpos{green}
\def\cellcolorneg{white}
\tikzset{pics/cell/.style n args={1}{code={%
  \pgfmathparse{round(#1)}\edef\prob{\pgfmathresult}
  \pgfmathparse{(\prob<50)?100:0}
  \colorlet{contentcolor}{\cellcolorpos!\pgfmathresult!\cellcolorneg}
  \fill[\cellcolorpos!\prob!\cellcolorneg]
    rectangle (\cellwidth,\cellheight) +(0,-\cellheight) coordinate (C);
  \path (C)+(0,0.1) node[contentcolor,anchor=base east]
    {\ifdim#1 pt>0pt #1\else{$\ast$\hspace*{1em}}\fi};
}}}

\title{Stop Simulating!\\Efficient Computation of\\Tournament Winning Probabilities}
\author{\mbox{Ulrik Brandes \quad Gordana Marmulla \quad Ivana Smokovic}\\
ETH Zürich, Social Networks Lab}
\date{Working Paper (\today)}

\begin{document}
\maketitle
\begin{abstract}
In the run-up to any major sports tournament, 
winning probabilities of participants
are publicized for engagement and betting purposes. 
These are generally based on simulating the tournament
tens of thousands of times by sampling from single-match outcome models.
We show that,
by virtue of the tournament schedule,
exact computation of winning probabilties
can be substantially faster than their approximation through simulation.
This notably applies to the 2022 and 2023 FIFA World Cup Finals,
and is independent of the model used for individual match outcomes.
\end{abstract}
\enlargethispage{3ex}

\section{Introduction}

Predicting sports tournament winning probabilities is not just a popular pastime
and an academic proxy competition, but also sets expectations
and thus informs bookmaker odds and gamblers' bet placements~\citep{winston_mathletics_2022}.
Hence, a flurry of predictions is published before each major tournament.
Our focus here is on association football (soccer),
and specifically the most recent FIFA World Cup Finals,
but the technique we introduce is more generally applicable.

Winning probabilities are commonly obtained via tournament simulations
sampling from single-match outcome prediction models.
While attention-seeking media tend to use cute animals to sample match outcomes
(see also \citealt{horvat_paul_2020} for their proper naming),
machine learning and other statistical models~\citep{tsokos_modeling_2019,horvat_use_2020}
enable repeated sampling to stabilize predictions.
Prominent examples of simulation-based predictions
include those of
Website \emph{FiveThirtyEight},%
\footnote{\url{https://projects.fivethirtyeight.com/2022-world-cup-predictions/}, accessed 19~July 2023}
the Alan Turing Institute,%
\footnote{\url{https://github.com/alan-turing-institute/WorldCupPrediction}, accessed 19~July 2023}
Joshua Bull,%
\footnote{\url{https://www.maths.ox.ac.uk/node/61756}, accessed 19~July 2023}
the \emph{DTAI Sports Analytics Lab},%
\footnote{\url{https://dtai.cs.kuleuven.be/sports/worldcup2023/}, accessed 19~July 2023}
and other groups of academic forecasters.%
\footnote{\url{https://www.zeileis.org/news/fifawomen2023/}, accessed 19~July 2023}

We do not attempt to add to the rich list of outcome prediction models,
but rather question whether a shared feature~--
the use tournament simulations to extend single-match predictions to winning probabilities~--
is computationally efficient.
Published probabilities would often be based on 100,000 simulation runs, 
and never less than 10,000.
Our main contribution is an algorithm that computes
the exact probabilities implied by a single-match model 
in a time equivalent to a few hundred simulation runs.

As has been discussed multiple times,
probabilities are efficiently computed in a bottom-up traversal of the tournament bracket,
if the seeding of teams is fixed~\citep{edwards_combinatorial_1991,schwertman_probability_1991,bettisworth_phylourny_2023}.
For a variety of reasons,
including a guaranteed minimum number of matches for each participating team,
seeding in the bracket is generally not fixed,
but determined in a preceding group phase.
For tournaments such as the most recent FIFA World Cup Finals
the number of possible bracket seedings is in the hundreds of millions 
and therefore prohibitively expensive to enumerate.
\citet{koning_simulation_2003} propose a hybrid approach 
in which they simulate only the group stage
and then calculate winning probabilities conditional on the sampled group ranking.
Our approach differs in that we enumerate outcomes for each group separately, 
and propagate the rank probabilities through the bracket
while considering dependencies created by teams on trajectories that meet,
especially those that already originate from the same group. 
By making these dependencies explicit
we are able to exploit independence among the remaining outcomes
for efficient computation of exact probabilities.

Our approach therefore relies on limited mixing of team trajectories
in the current world cup formats.
There are several other tournament formats
that allow for the computation of exact probabilities.
Among them are single-elimination tournaments 
with random seeding~\citep{david_tournaments_1959,hartigan_probabilistic_1966},
double-elimination tournaments 
with a given seeding~\citep{edwards_double-elimination_1996},
a variant of single- and double-elimination in which
a team must win two matches against an opponent to advance~\citep{searls_probability_1963},
and a random knockout tournament for the case
that the number of participants is not a power of two~\citep{narayana_contributions_1969}.

In addition to the scenarios above,
exact winning probabilities can further be used
to assess seeding criteria empirically~\citep{david_method_1988}.
The issue of effectiveness, i.e., whether the highest rated team is most likely to win,
has been studied for various tournament designs~\citep{glenn_comparison_1960,maurer_most_1975,hwang_several_1977,chung_stronger_1978,appleton_may_1995,mcgarry_efficacy_1997,schwenk_what_2000,marchand_comparison_2002}.
This includes, in particular,
FIFA World Cups~\citep{scarf_numerical_2011,cea_analytics_2020,sziklai_efficacy_2022}.

The remainder of this paper is organized as follows.
In Section~\ref{sec:preliminaries},
we define terminology and the tournament format considered.
An examplary single-match outcome model is introduced in Section~\ref{sec:oracle},
and the extension of any such model to tournament winning probabilities
is described in Section~\ref{sec:algorithm}.
Results for the~2022 and~2023 FIFA World Cups are presented in Section~\ref{sec:results}
before we conclude in Section~\ref{sec:conclusion}.

\section{Preliminaries}\label{sec:preliminaries}

We consider tournaments in which $n$~teams compete for a single title.
The goal is to determine the probability of each team to win that title, 
given an oracle that returns outcome probabilities for matches between any two teams.
A number of other results are obtained as a byproduct,
for instance the probabilities of teams exiting
the tournament in a particular round.

\subsection{Tournament schedule and notation}

\begin{figure}
\begin{subfigure}{\linewidth}
  \includegraphics[width=\linewidth]{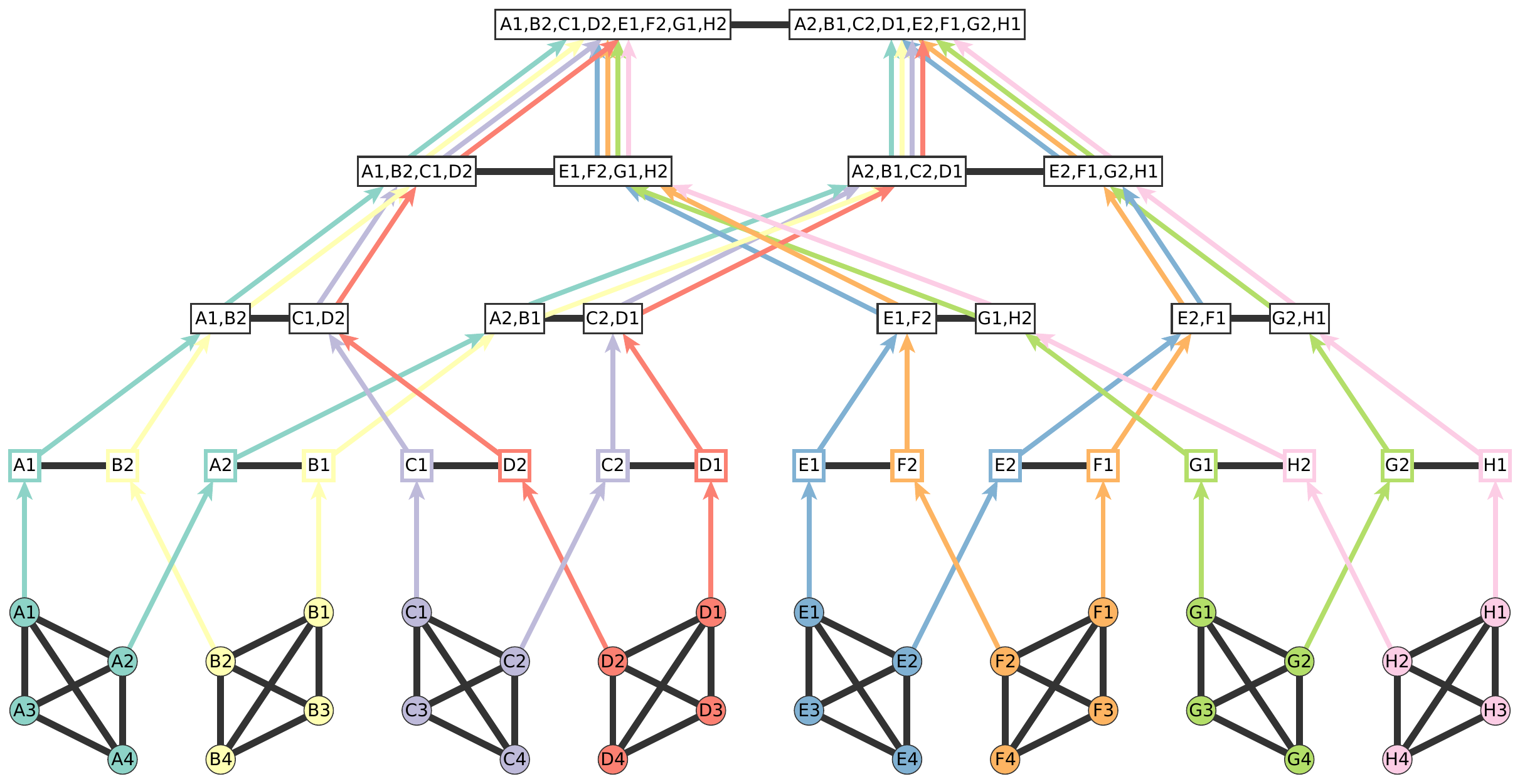}
  \caption{FIFA Men's World Cup 2022}
\end{subfigure}
\begin{subfigure}{\linewidth}
  \bigskip
  \includegraphics[width=\linewidth]{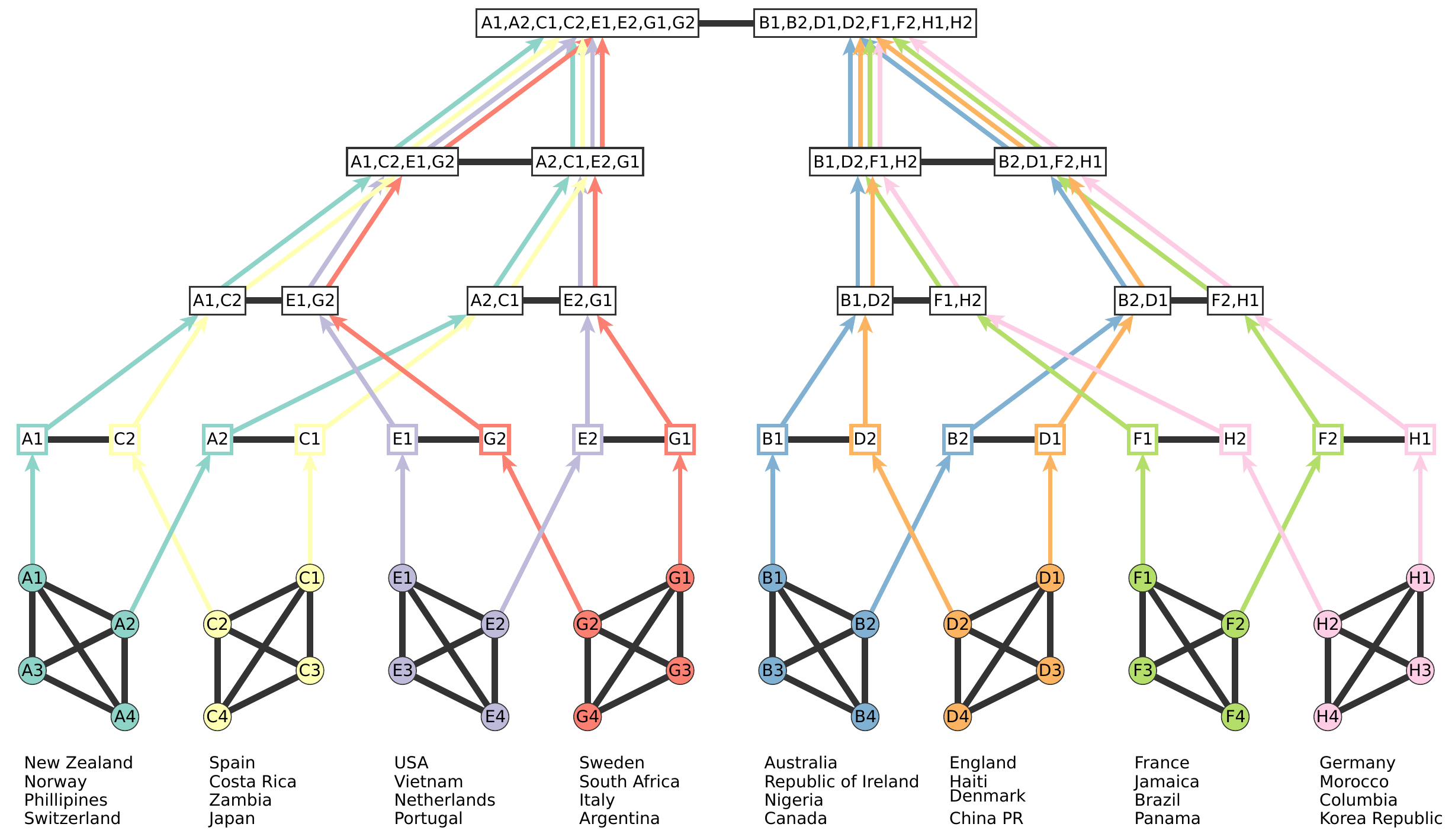}
  \caption{FIFA Women's World Cup 2023}
\end{subfigure}
\caption{Tournament schedules for $32$~teams in $8$~groups.}
\label{fig:schedule}
\end{figure}

Denote by $N=\{0,\ldots,n-1\}$ the set of participating teams.
While there are many ways to design a sports tournament~\citep{scarf_numerical_2009},
we are specifically interested in
the FIFA Men's World Cup~2022 and FIFA Women's World Cup~2023,
where $n=32$ in both cases.
In their rather common tournament formats,
the schedule is fixed in advance and consists of two phases:
\begin{enumerate}
\item \emph{Group stage:}
  participants compete in $m=8$~equal-sized groups $N_0,\ldots,N_{m-1}$
  to which they have been assigned in a random draw from $n/m=4$~urns (``pots'')
  such that every group consists of one team from each pot.%
  \footnote{Pots are generally stratified with respect to past performance to balance expected opponent strengths~\citep{engist_effect_2021}.}
  W.l.o.g., we assume $N_g=\{g\cdot\frac{n}{m},\ldots,(g+1)\cdot\frac{n}{m}-1\}$,
  $g=0,\ldots,m-1$,
  so that team~$i$ is in group $N_{i\div\frac{n}{m}}$,
  where $\div$ denotes integer division without remainder (e.g., $9\div 4=11\div 4=2$).
  After separate round-robin tournaments
  in which each team plays every other team from the same group,
  the two teams ranking first and second in each group advance to the elimination rounds,
  and the other $16$~teams are eliminated from the competition.
\item \emph{Knockout stage:}
  the remaining $2m=16$~teams are paired in $\log_2(m)=4$~successive rounds of elimination matches,
  each time halving the number of teams remaining in competition for the title.
\end{enumerate}
As shown in Figure~\ref{fig:schedule},
the fixtures of the two focal tournaments differ
in the way teams can meet in the elimination rounds.
We will see below that this matters also for computation time,
because the mixing of team trajectories
creates dependencies among winning probabilities.

\subsection{Assumptions}

We assume knowledge of the outcome distribution of individual matches
on the level of wins, draws, and losses,
i.e., we do not consider scorelines.
The basis on which match outcome probabilities are determined varies.
Proposed models often include past performance, but also
player roster and market value,
match location and weather conditions, 
betting behavior and social media signals,
and for international sides even
country characteristics and stock market trends~\citep{hubacek_forty_2022,groll_prediction_2015,schauberger_predicting_2018,kuper_soccernomics_2022,batarfi_why_2021,lepschy_success_2020}.

Our computational scheme is independent of
the particular model used to predict match outcomes.
We only require the following conditions to be met:
\begin{itemize}
\item predictions are to be made before the start of the tournament, 
i.e., in particular without evidence from matches and developments during the tournament
\item match outcome probabilities are independent,
i.e., based only on information about the two teams playing each other
\item the only outcomes distinguished are winning, drawing, and losing,
i.e., tie-breaking to determine a ranking at the end of the group stage
(based on goal differences, direct comparisons, fair-play etc.)
as well as extra time and penalty shootouts in the elimination phase 
are not considered and instead treated as fair coin flips.
\end{itemize}
While most of these assumptions could be relaxed or dropped altogether,
they are instructive and simplify the exposition.

\section{Single-Match Outcome Probabilities}\label{sec:oracle}

With the assumptions above, we can pre-compute match outcome distributions
for an upper-triangular matrix of random variables
$\left(M_{ij}\right)_{i<j\in N}$ with values in $\{0,1,2\}$
for a loss, draw, or win of team~$i$ when playing against team~$j$.
The diagonal is empty (teams are not playing themselves)
and the other off-diagonal entries are implicit,
because transposed outcomes are related by $M_{ji}=2-M_{ij}$.

This ternary representation will prove more convenient
than a symmetric $\{-1,0,1\}$ or point-based $\{0,1,3\}$ representation,
because it allows us to encode sequences of outcomes as single integers
with ternary digits.

Elimination matches end with one team advancing.
For ease of exposition, we split the probability $P(M_{ij}=1)$ of a draw
evenly between the probabilities of winning and losing.
Only two possible outcomes remain and by setting
the probability of~$i$ advancing by eliminating~$j$ to
$$
  P(M'_{ij}=1)=\frac{1}{2}\cdot P(M_{ij}=1)+P(M_{ij}=2)
$$
we also obtain $P(M'_{ij}=0)=1-P(M'_{ij}=1)=\frac{1}{2}\cdot P(M_{ij}=1)+P(M_{ij}=0)$.
In other words, we make no predictions about one team having higher chances
in extra time or in penalty shootouts, the standard mechanisms to break a tie. 
This could be included, of course. 

In principle, the algorithm described in the next section
can be applied to any model for single-match outcomes.
For demonstration purposes we choose here a model that is simple,
yet relevant, as it underlies the FIFA/Coca-Cola World Rankings~\citep{szczecinski_fifa_2022}.
These rankings are variants of \Elo ratings~\citep{elo_rating_2008} 
and therefore based on a point system that rewards past results.
Teams are assigned points $\rho_i$, $i=0,\ldots,n-1$,
which are updated regularly based on
the differences between expected and actual match outcomes.
Expected outcomes are determined by a logistic function in the differences of points,
$$
   0<\frac{1}{1 + 10^{(\rho_j - \rho_i)/\sigma}}<1
$$
so that equally rated teams~$i,j$ yield an expectation of $\frac{1}{2}$.
The rate at which the expectation tends to~$0$ for $\rho_i<\rho_j$,
and to~$1$ for $\rho_i>\rho_j$,
is governed by sensitivity scale~$\sigma$.
The update mechanisms,
and in fact even the way an outcome is determined,
differ between the FIFA rankings for men and women.
Except for the scaling factor, however, expected outcomes are determined the same. 
FIFA uses $\sigma=600$ in the men's ranking,
and while the scale is not published for the women's ranking,
the magnitudes of recent updates suggest that it is $\sigma=400$.

To turn the continuous value for expected match outcomes
into a discrete distribution over losses, draws (group stage only), and wins, 
we follow the model of \citet{bradley_rank_1952}.
For a match between teams~$i,j$,
there are two perspectives on the expected outcome,
$$
\theta_i = \frac{1}{1 + 10^{(\rho_j - \rho_i)/\sigma}}
\qquad\qquad
\theta_j = \frac{1}{1 + 10^{(\rho_i - \rho_j)/\sigma}} = 1-\theta_i
$$
which we interpret as the respective team strengths.
Draws during the group stage are accomodated by
the extension of \citet{davidson_extending_1977}.
$$\renewcommand\arraystretch{1.7}
  \begin{array}[t]{c|c|c}
  \multicolumn{3}{c}{\textbf{group stage}}\\[1ex]
  P(M_{ij}=0) & P(M_{ij}=1) & P(M_{ij}=2)\\\hline
  \frac{\theta_j}{\theta_i + \sqrt{\theta_i\theta_j} + \theta_j} &
  \frac{\sqrt{\theta_i\theta_j}}{\theta_i + \sqrt{\theta_i\theta_j} + \theta_j} &
  \frac{\theta_i}{\theta_i + \sqrt{\theta_i\theta_j} + \theta_j} \\[1ex]\hline
  \text{$i$ lose} & \text{$i$ and $j$ draw} & \text{$i$ win}
  \end{array}
  \qquad\qquad
  \begin{array}[t]{c|c}
  \multicolumn{2}{c}{\textbf{knockout stage}}\\[1ex]
  P(M'_{ij}=0) & P(M'_{ij}=1)\\\hline
  \frac{\theta_j}{\theta_i + \theta_j} &
  \frac{\theta_i}{\theta_i + \theta_j} \\[1ex]\hline
  \text{$i$ exit} & \text{$i$ advance}
  \end{array}
$$
In a final step we re-calibrate the model by adjusting scaling factor~$\sigma$. 
Since~$\sigma$ has an influence on the number of points re-allocated between teams,
it appears that the above values are motivated in part by an attempt to dampen fluctuations,
and as such too conservative for predictions in a high-stakes competition.
We therefore determined the scaling factor for the women's world cup
by computing tournament winning probabilities (as described in the next section)
for different scaling factors, and then chose $\sigma=240$ as the minimizer of
the squared differences between winning probabilities and averaged betting odds
before the start of the tournament.
Since this is $60\%$ of the estimated value of $400$ used in the FIFA ranking,
we made the same adjustment for the men's by setting $\sigma=0.6\cdot600=360$.

\begin{figure}[htb]
\begin{subfigure}{0.5\linewidth}
  \includegraphics[width=\linewidth]{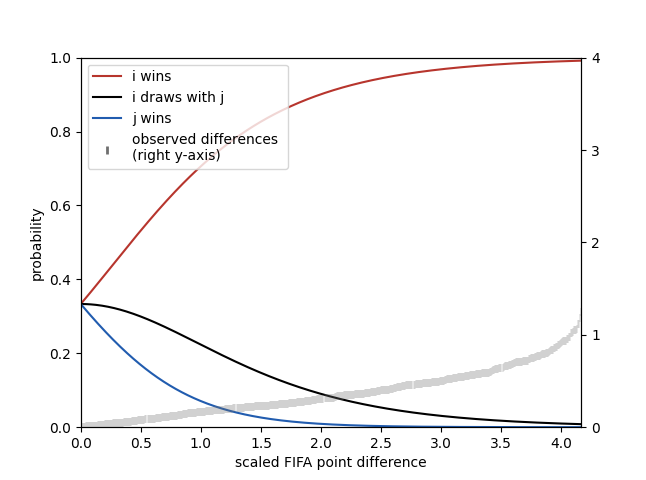}
    \caption{FIFA Men's World Cup 2022}
\end{subfigure}
\begin{subfigure}{0.5\linewidth}
  \includegraphics[width=\linewidth]{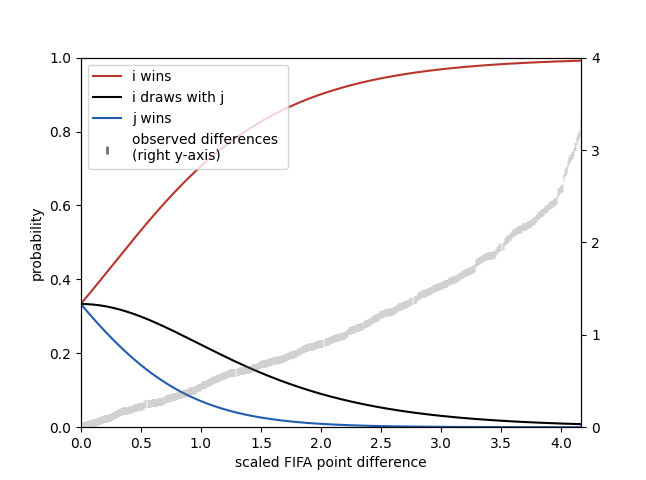}
    \caption{FIFA Women's World Cup 2023}
\end{subfigure}
\caption{Modeled outcome probabilities
as a function of re-scaled absolute FIFA point differences $|\rho_i-\rho_j|/\sigma$.
Actually occurring values for all pairs of teams are shown in non-decreasing order.
}
\label{fig:oracle}
\end{figure}

As illustrated in Figure~\ref{fig:oracle},
estimated team strengths not only differ more strongly in the women's tournament,
but are also more skewed.
Despite the different scaling factors,
this generally leads to more definite predictions in the women's case.

\section{Tournament Probabilities}\label{sec:algorithm}

It is important to note that in the tournaments we consider
all fixtures are determined in advance.
The possible trajectory of a team is fixed
as soon as their table rank is known at the end of the group stage.
In contrast, most national and international cup competitions for clubs
feature additional draws after group stages and elimination rounds.

\subsection{Group stage}

The group stage is organized into separate round-robin tournaments within each group.
Whether a team advances through this stage thus
depends on the results of all matches in the group,
including those between other teams,
but not on matches in other groups. 

We assume that match outcomes are independent from each other,
both within and across groups.
This assumptions is violated, for instance,
if a team attempts to engineer a result
to prevent a specific team in their group from entering the knockout stage,
or to avoid a specific opponent from another group by entering it in a different seed position.

To determine the probability of ranks in the final group standings,
all possible combinations of match outcomes are enumerated in each group.
Hardness results for round-robin tournaments suggest that
there is no substantially more efficient alternative~\citep{bernholt_football_1999,kern_new_2001,bernholt_komplexitatstheorie_2002}.

There are $4$~teams per group, playing a total of $\binom{4}{2}=6$ matches,
which leads to $3^6=729$ possible combinations of outcomes.
Because of the independence assumption,%
\footnote{Since all combinations are considered anyway,
it would be possible, for instance,
to make later results depend on earlier ones.}
the joint probability of each combination
is given by the product of its six single-match outcome probabilities.

Moreover, all these outcome combinations are disjoint events, 
so that the probability of a team finishing in a specific table position
is the sum of the probabilities of all six-match combinations
that lead to this position. 
If an outcome combination leads to two or more teams having the same number of points,
we rank them uniformly at random,
because we did not require single-match outcome distributions
to include probabilities for exact scorelines,
and tie-breaking rules may depend on even more factors.

To enumerate all outcome sequences,
we associate them with integers $0\leq s<729$.
In their ternary representation, $s=s_5s_4s_3s_2s_1s_0\in\{0,1,2\}^6$,
we let digit $s_d\in\{0,1,2\}$ correspond to the result of the $d$th match, $d=0,\ldots,5$.
As in the previous section, $0,1,2$ indicate a loss, draw, or win.
We then number matches in relation to the pots from which teams were drawn into the group,
and order them column-wise for ease of index calculation in Algorithm~\ref{alg:group}.
$$
\begin{array}{r|cccc}
\text{outcomes} & 0 & 1 & 2 & 3\\\hline
0 & \cdot & s_0 & s_1 & s_3\\
1 & 2-s_0 & \cdot & s_2 & s_4\\
2 & 2-s_1 & 2-s_2 & \cdot & s_5\\
3 & 2-s_3 & 2-s_4 & 2-s_5 & \cdot
\end{array}
$$
Any given outcome sequence determines the number of points for each team
and thus the final group standings separately in each group.
Since we break ties at random, the same outcome sequence
can lead to different rankings and, in particular,
different teams advancing.
We therefore associate with each outcome sequence a list of pairs
containing all possible combinations of teams advancing
as first and second into the round of the last~16.
Only $1260$ such pairs actually occur across all outcome sequences,
so that iterating over these lists is more efficient
than iterating over the rows of a table with all 
$729\times12=8748$ combinations of pairs and outcome sequences.%
\footnote{To simplify the management of data structures,
we can nevertheless store the lists in a $729\times13$ matrix of integers,
in which the entry in the first column of row~$s$
gives the number~$1\leq q\leq12$ of possible top-ranked pairs for outcome sequence~$s$,
and integer entries~$0\leq t\leq15$ in the next~$q$ columns represent the pair $(t\div4,t\mod4)$.}

\begin{figure}[tb]
$\begin{array}[t]{rl}
\multicolumn{2}{c}{\text{\bf sequence}}\\[1ex]
s & =279\\
  & =201100_3\\
\end{array}$
\hfill
$\begin{array}[t]{r|cccc}
\multicolumn{5}{c}{\text{\bf outcomes}}\\[1ex]
  & 0 & 1 & 2 & 3\\\hline
0 & \cdot & 0 & 0 & 1\\
1 &   & \cdot & 1 & 0\\
2 &   &   & \cdot & 2\\
3 &   &   &   & \cdot
\end{array}$
\hfill
\begin{tabular}[t]{r|cc}
\multicolumn{3}{c}{\bf group table}\\[1ex]
rank & team & points \\\hline
1. & 2 & 7 \\\hline
2. & 1 & 4 \\
2. & 3 & 4 \\\hline
4. & 0 & 1 \\
\end{tabular}
\hfill
$\begin{array}[t]{rl}
\multicolumn{2}{c}{\text{\bf possible rankings}}\\[1ex]
R_{279}: & \{(2,1),\,(2,3)\}
\end{array}$
\caption{An integer representing a sequence of match outcomes
in ternary representation (loss/draw/win)
yields a table in which two teams are tied for second place by points.
Therefore, team~2 advances as the group's winner,
and with equal probability either team~1 or~3 in second place. 
No other pair can advance with this sequence of match outcomes.}
\label{fig:groupoutcome}
\end{figure}

\begin{algorithm}[tb]
\caption{\textbf{Group stage.}
Determining probabilities $G_{ij}$ of teams $i,j\in\{0,\ldots,31\}$
to jointly advance from the same group in first and second place}
$G\gets0^{32\times 32}$\;
\For{$s=0,\ldots,728$}
 {
  \lFor{$d=0,\ldots,5$}{$s_d\gets(s\div3^d)\mod3$}
  \For{$g=0,\dots,7$}
   {
    $p\gets \prod_{i<j\in\{0,\ldots,3\}} M[4g+i,4g+j,s_{\binom{j}{2}+i}]$\;
    \lForEach{$(i_1,i_2)\in R[s]$}
     {increase $G[4g+i_1,4g+i_2]$ by $p\cdot\frac{1}{|R[s]|}$}
   }
 }
\label{alg:group}
\end{algorithm}

Algorithm~\ref{alg:group} summarizes
the computation of probabilities $G_{ij}$ for all pairs of teams
to jointly advance in first and second place from the group stage.
The $729$ outcomes are enumerated and for every group 
their respective probability~$p$ is determined
as the product of independent match outcome probabilities. 
For each ordered pair of teams in a group, 
the probability to advance together is obtained 
as the expected value of the relative frequency
with which they finish in the first two places in this order.
In other words, the probability of the potentially advancing pair is increased by
the probability of said outcome sequence
weighted by the inverse of the number of different pairs in the first two places
when ties are broken at random.
As a byproduct we obtain $1-\sum_{j} G_{ij}+G_{ji}$
as the probability that team~$i\in\{0,\ldots,31\}$
exits the tournament after the group stage.

Computation time is dominated by the enumeration of all outcome sequences,
and their evaluation in each group.
Entries in the pre-computed lists~$R$ of possibly top-ranked pairs
are iterated over once per group,
so that the amortized number of elementary steps is $8\times(729+1260)=15,912$.

We note that the result matrix~$G$ has at most $8\times4\times3=96$ non-zero entries,
and could clearly be stored more space-efficiently.
We use it here only to highlight similarity with, and possible re-use in, the other rounds.


\subsection{Knockout stage}

With half of the $32$~teams eliminated after the group stage, 
the knockout stage consists of 
$8$~matches in the round of the last~$16$,
$4$~quarterfinals, $2$~semifinals, and $1$~final.

Curiously, the schedule differs between the two FIFA tournaments
as shown in Figure~\ref{fig:schedule}.
While it was possible for two teams from the same group
to play the final in the men's competition,
(had they won their semifinals, Croatia and Morocco, both from Group~F,
would have played the final rather then the match for 3rd place)
this is not possible in the women's competition,
because the tournament bracket separates the eight groups
into two subtrees of four 
that do not mix before the final. 
As a consequence of this difference, for instance,
England (Group~D) and the USA (Group~E) 
will not have to face each other before the final, if at all.

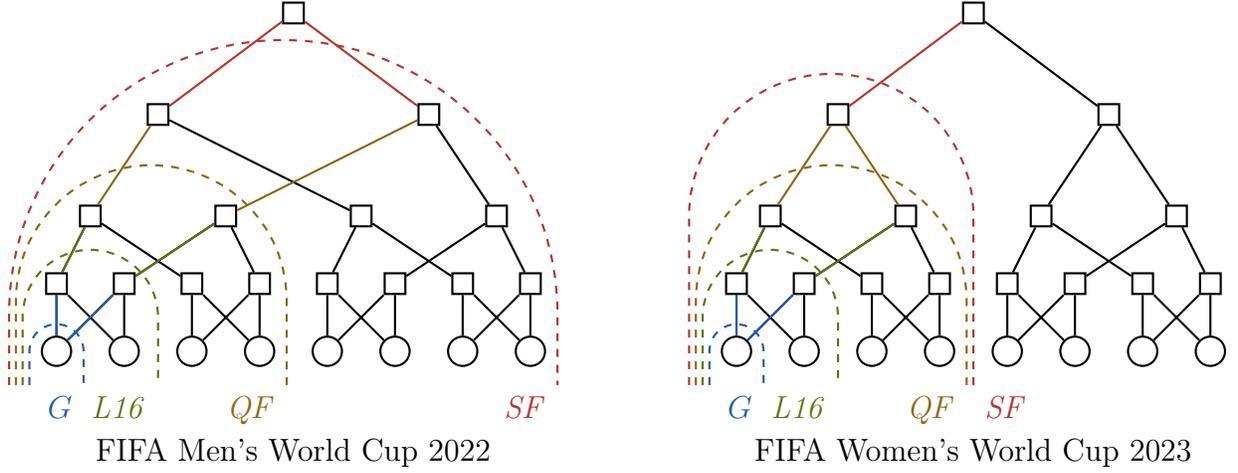
\begin{figure}[tb]
\newcommand{\bracket}{\foreach \g in {1,...,8}
 { \node[draw,circle] (G\g) at (2*\g,0) {};
   \node[draw,rectangle] (L\g) at (2*\g,2) {};
 }
\foreach \g/\h in {1/2,3/4,5/6,7/8}
 { \draw (G\g) -- (L\g) -- (G\h) -- (L\h) -- (G\g); }
\foreach \g [count=\n] in {3,7,11,15} { \node[draw,rectangle] (Q\n) at (\g,4) {}; }
\foreach \g/\h [count=\n] in {1/3,2/4,5/7,6/8}
 { \draw (L\g) -- (Q\n) -- (L\h); }
\node[draw,rectangle] (S1) at (5,7) {};
\node[draw,rectangle] (S2) at (13,7) {};
\node[draw,rectangle] (F) at (9,10) {};
\draw[blue] (L1) -- (G1) -- (L2);
\draw[dashed,blue]
  (1.2,-1) -- +(0,1) arc (180:0:0.8) -- +(0,-1) node[below left]{$G$};
\draw[green] (L1) -- (Q1) (L2) -- (Q2);
\draw[dashed,green]
  (1,-1) -- +(0,2) arc (180:0:2) -- +(0,-2) node[below left]{$\eighthfinal$};
\draw[dashed,yellow]
  (0.8,-1) -- +(0,2.5) arc (180:0:4) -- +(0,-2.5) node[below left]{$\quarterfinal$};
}
\begin{tikzpicture}[scale=0.45,thick]
\bracket
\draw (Q3) -- (S1) (Q4) -- (S2);
\draw[yellow] (Q1) -- (S1) (Q2) -- (S2);
\draw[dashed,red]
  (0.6,-1) -- ++(0,2.1) arc (180:0:8.1) -- ++(0,-2.1) node[below left]{$\semifinal$};
\draw[red] (S1) -- (F) -- (S2);
\draw (9,-3) node{FIFA Men's World Cup 2022};
\end{tikzpicture}
\hfill
\begin{tikzpicture}[scale=0.45,thick]
\bracket
\draw (Q3) -- (S2) -- (Q4) (S2) -- (F);
\draw[yellow] (Q1) -- (S1) -- (Q2);
\draw[dashed,red]
  (0.6,-1) -- +(0,5) arc (180:0:4.2) -- +(0,-5) node[below right]{$\semifinal$};
\draw[red] (S1) -- (F);
\draw (9,-3) node{FIFA Women's World Cup 2023};
\end{tikzpicture}
\caption{The probabilities of teams advancing to the next round
(crossing a dashed line) 
are dependent only for teams from the same set of groups
(inside the dashed line).}
\label{fig:rainbows}
\end{figure}

In both tournaments, the brackets are set up such that
each group sends a pair of teams into the round of the last~16
where they face a pair of opponents from the same other group,
i.e., their two subtrees are mixing.
The key observation,
depicted in Figure~\ref{fig:rainbows}
is that whatever happens in these two matches
is independent from all other groups,
and that this continues up the elimination tree. 
In each round, there is pairwise merging of intervals of consecutive groups.

In the first elimination round, the round of the last~16,
a team can face any of the $4$~teams from a specific other group,
while another pair of teams from the same two groups faces each other.
The probability of these pairs of matches happening are not independent,
because the probability of a team finishing first in the group
influences the probabilities that some team finishes second,
in particular it reduces said probability to zero for the team itself.

\begin{algorithm}[tb]
\caption{\textbf{Round of Last~16.}
Determining probabilities $\eighthfinal_{ij}$ of teams $i,j$
from mixing subtrees to advance jointly into the quarterfinals.}
$\eighthfinal\gets0^{32\times32}$\;
\For{$g=0,2,4,6$}
 {
  \For{$i\neq j\in\{4g,\ldots,4g+3\}$}
   {
    \For{$k\neq\ell\in\{4(g+1),\ldots,4(g+1)+3\}$} 
     { 
      $p\gets G[i,j]\cdot G[k,\ell]$\;
      increase $\eighthfinal[i,j]$ by $p\cdot M'[i,\ell]\cdot M'[j,k])$\;
      increase $\eighthfinal[i,k]$ by $p\cdot M'[i,\ell]\cdot M'[k,j])$\;
      increase $\eighthfinal[\ell,j]$ by $p\cdot M'[\ell,i]\cdot M'[j,k])$\;
      increase $\eighthfinal[\ell,k]$ by $p\cdot M'[\ell,i]\cdot M'[k,j])$\;
     }
   }
 } 
\label{alg:eighthfinal}
\end{algorithm}

As for the groups, we therefore compute probabilities to advance
for pairs of teams that win the two matches in which two subtrees are mixing.
Possible fixtures are obtained by combining
all pairs of teams advancing from one subtree
with all those advancing from another. 

The round of the last~16 (Algorithm~\ref{alg:eighthfinal})
and the quarterfinals (Algorithm~\ref{alg:quarterfinal})
can be implemented using the same principles.
Because this part of the schedule is shared, they apply to both world cups.
The number of elementary steps is determined by the number of matches
and the number of pairs that can populate each of them.
In $4$~pairs of matches in the round of the last~16,
$12$~pairs from one group are matched (cross-wise, first against second)
with $12$~pairs from another group for $4\times12^2=576$ cases to consider per match.
There are only $2$~pairs of quarterfinals,
but the number of ordered pairs from two already mixed groups
grows to $8\times7$ for a total of $2\times56^2=6272$ possible matchups.
Note that the indexing in Algorithm~\ref{alg:quarterfinal} is different,
because incoming pairs are no longer ordered as first and second in group,
but by trajectory from left to right.

\begin{algorithm}[tb]
\caption{\textbf{Quarterfinals.}
Determining probabilities $\quarterfinal_{ij}$ of teams $i,j$
from mixing subtrees to advance jointly into the semifinals.}
$\quarterfinal\gets0^{32\times32}$\;
\For{$g=0,4$}
 {
  \For{$i\neq j\in\{4g,\ldots,4g+7\}$}
   {
    \For{$k\neq\ell\in\{4(g+2),\ldots,4(g+2)+7\}$} 
     { 
      $p\gets\eighthfinal[i,j]\cdot\eighthfinal[k,\ell]$\;
      increase $\quarterfinal[i,j]$ by $p\cdot M'[i,k]\cdot M'[j,\ell])$\;
      increase $\quarterfinal[i,\ell]$ by $p\cdot M'[i,k]\cdot M'[\ell,j])$\;
      increase $\quarterfinal[k,j]$ by $p\cdot M'[k,i]\cdot M'[j,\ell])$\;
      increase $\quarterfinal[k,\ell]$ by $p\cdot M'[k,i]\cdot M'[\ell,j])$\;
     }
   }
 } 
\label{alg:quarterfinal}
\end{algorithm}

\begin{algorithm}[tb]
\caption{\textbf{Semifinals.}
Determining probabilities $\semifinal_{ij}$ of teams $i,j$
from mixing subtrees (men's)
or probabilities $\semifinal_i$ of teams~$i$
from the same subtree (women's)
to advance into the final.}
\begin{tabular}{@{}p{0.5\linewidth}p{0.4\linewidth}}
\bf World Cup 2022: & \bf World Cup 2023:\\[1ex]
$\semifinal\gets0^{32\times32}$\;
\For{$i\neq j\in\{0,\ldots,15\}$}
 {
  \For{$k\neq\ell\in\{16,\ldots,31\}$} 
   { 
    $p\gets\quarterfinal[i,j]\cdot\quarterfinal[k,\ell]$\;
    increase $\semifinal[i,j]$ by $p\cdot M'[i,k]\cdot M'[j,\ell])$\;
    increase $\semifinal[i,k]$ by $p\cdot M'[i,k]\cdot M'[\ell,j])$\;
    increase $\semifinal[\ell,j]$ by $p\cdot M'[k,i]\cdot M'[j,\ell])$\;
    increase $\semifinal[\ell,k]$ by $p\cdot M'[k,i]\cdot M'[\ell,j])$\;
   }
 } 
&
$\semifinal\gets0^{32\times1}$\;
\For{$g=0,4$}
 {
  \For{$i\neq j\in\{4g,\ldots,4g+15\}$} 
   { 
    $p\gets\quarterfinal[i,j]$\;
    increase $\semifinal[i]$ by $p\cdot M'[i,j])$\;
    increase $\semifinal[j]$ by $p\cdot M'[j,i])$\;
   }
 } 
\end{tabular}
\label{alg:semifinal}
\end{algorithm}

In the men's competition this scheme continued with the two semifinals
where pairs of teams from four mixed groups
face pairs of teams from the other four already mixed groups.
Since there are $16\times15=240$ possible pairs of teams advancing from four groups,
we have a total of $240^2=57,600$ cases to consider.
In the women's competition, however,
the semifinals are played between the two pairs
of teams advancing together from the quarterfinals,
so that there are only $2\times240=480$ possible matchups.
The pseudo-code in Algorithm~\ref{alg:semifinal} reflects
this difference between the fixtures from Figure~\ref{fig:rainbows}.

\begin{algorithm}[tb]
\caption{\textbf{Final.}
Determining probability $F_i$ of team $i$ to win the final.}
$F\gets0^{32\times1}$\;
\begin{tabular}{@{}p{0.4\linewidth}p{0.5\linewidth}}
\bf World Cup 2022: & \bf World Cup 2023:\\[1ex]
\For{$i\neq j\in\{0,\ldots,31\}$}
 {
  $p\gets\semifinal[i,j]$\;
  increase $F_i$ by $p\cdot M'[i,j]$\;
  increase $F_j$ by $p\cdot M'[j,i]$\;
 }
&
 \For{$(i,j)\in\{0,\ldots,15\}\times\{16,\ldots,31\}$}
 {
  $p\gets\semifinal[i]\cdot\semifinal[j]$\;
  increase $F_i$ by $p\cdot M'[i,j]$\;
  increase $F_j$ by $p\cdot M'[j,i]$\;
 }
\end{tabular}
\label{alg:final}
\end{algorithm}

Consequently, every team could meet any other team in the final of the 2022~World Cup,
and even on two different routes initiated by finishing in reverse order in their groups.
In the schedule for the 2023~World Cup, only teams from separate subtrees can play the final.
The enumerations differ therefore, again, in Algorithm~\ref{alg:final},
and yield $32\times31=992$ and $16^2=256$ cases, respectively.
\enlargethispage{-6ex}

The aggregate number of cases considered is therefore given by the following account:
\begin{center}
  \begin{tabular}{r|rr}
  combinations  & \bf 2022 & \bf 2023\\\hline
  group stage   & 15,912   & 15,912\\
  last 16       & 576      & 576\\
  quarterfinals & 6,272    & 6,272\\ 
  semifinals    & 56,700   & 240\\
  final         & 992      & 256\\\hline
  \bf total & \bf 80,452   & \bf 23,256
  \end{tabular}
\end{center}
There are no large hidden constants in this computation of exact probabilities.
Any one simulation run, on the other hand,
samples $63$~match outcomes (not counting the match for third place). 
This suggests,
and is confirmed by computational experiments presented in Section~\ref{sec:results},
that the exact computation requires no more time
than is needed for a few hundred simulation runs.

\begin{figure}[tb]
\centering
\input{exact_probs_m}
\begin{tikzpicture}[transform shape]\sf
  \foreach \t/\w/\f/\s/\q/\l/\g [count=\n] in \probabilities
    {
      \draw (0,-\n*\cellheight) +(0,0.2) node[anchor=base west]{\t}
       ++(1.5,0)
       ++(\cellwidth,0) pic{cell={\w}}
       ++(\cellwidth,0) pic{cell={\f}}
       ++(\cellwidth,0) pic{cell={\s}}
       ++(\cellwidth,0) pic{cell={\q}}
       ++(\cellwidth,0) pic{cell={\l}};
    }
  \draw (1.5+\cellwidth,0.2)
    ++(\cellwidth,0) node[above,anchor=base east]{winner}
    ++(\cellwidth,0) node[above,anchor=base east]{final}
    ++(\cellwidth,0) node[above,anchor=base east]{semi}
    ++(\cellwidth,0) node[above,anchor=base east]{quarter}
    ++(\cellwidth,0) node[above,anchor=base east]{last 16};
\end{tikzpicture}
\caption{FIFA Men's World Cup 2022 probabilities (given as percentages)
to reach elimination rounds
based on points from FIFA/Coca-Cola World Ranking of October~2022.}
\label{fig:mwc2022-fifa}
\end{figure}

\begin{figure}[tb]
\centering
\input{exact_probs_w}
\begin{tikzpicture}[transform shape]\sf
  \foreach \t/\w/\f/\s/\q/\l/\g [count=\n] in \probabilities
    {
      \draw (0,-\n*\cellheight) +(0,0.2) node[anchor=base west]{\t}
       ++(1.5,0)
       ++(\cellwidth,0) pic{cell={\w}}
       ++(\cellwidth,0) pic{cell={\f}}
       ++(\cellwidth,0) pic{cell={\s}}
       ++(\cellwidth,0) pic{cell={\q}}
       ++(\cellwidth,0) pic{cell={\l}};
    }
  \draw (1.5+\cellwidth,0.2)
    ++(\cellwidth,0) node[above,anchor=base east]{winner}
    ++(\cellwidth,0) node[above,anchor=base east]{final}
    ++(\cellwidth,0) node[above,anchor=base east]{semi}
    ++(\cellwidth,0) node[above,anchor=base east]{quarter}
    ++(\cellwidth,0) node[above,anchor=base east]{last 16};
\end{tikzpicture}
\caption{FIFA Women's World Cup 2023 probabilities (given as percentages)
to reach elimination rounds
based on points from FIFA/Coca-Cola World Ranking of June~2023.
Percentages shown as~$\ast$ are greater than~$0$ but less than $0.00005$.}
\label{fig:wwc2023-fifa}
\end{figure}

\clearpage
\section{Results}\label{sec:results}

In this section we present winning probabilities
obtained by running our algorithm on top of
the single-match model introduced in Section~\ref{sec:oracle}.
Recall that other models will lead to other outcomes,
and that our objective is only to demonstrate computational feasibility,
not quality of prediction. 

Retrospective winning probabilities for the 2022~World Cup
are given in Figure~\ref{fig:mwc2022-fifa}.
Not surprisingly, the teams many consider to have deviated from prior expectations 
are semifinalists Morocco, and group-stage exits Belgium and Germany.
Belgium may also be the team that had the highest variance in predictions across models.

Winning probabilities for the 2023~World Cup
are given in Figure~\ref{fig:wwc2023-fifa}.
Due to a stronger skew in points in the women's ranking,
probabilities also differ more strongly between the top and bottom teams.
One consequence of the schedule is that
both France and Brazil are expected to advance from Group~F,
but in uncertain order. 
The trajectory of the team placed second leads past the winners of Groups~H and~D,
who are most likely Germany and England.
With one very difficult path,
their chances of advancing to the final are therefore reduced more strongly than, 
for instance, Australia's, who are expected to face Denmark or China from Group~D.

\begin{figure}[htb]
\includegraphics[width=\linewidth]{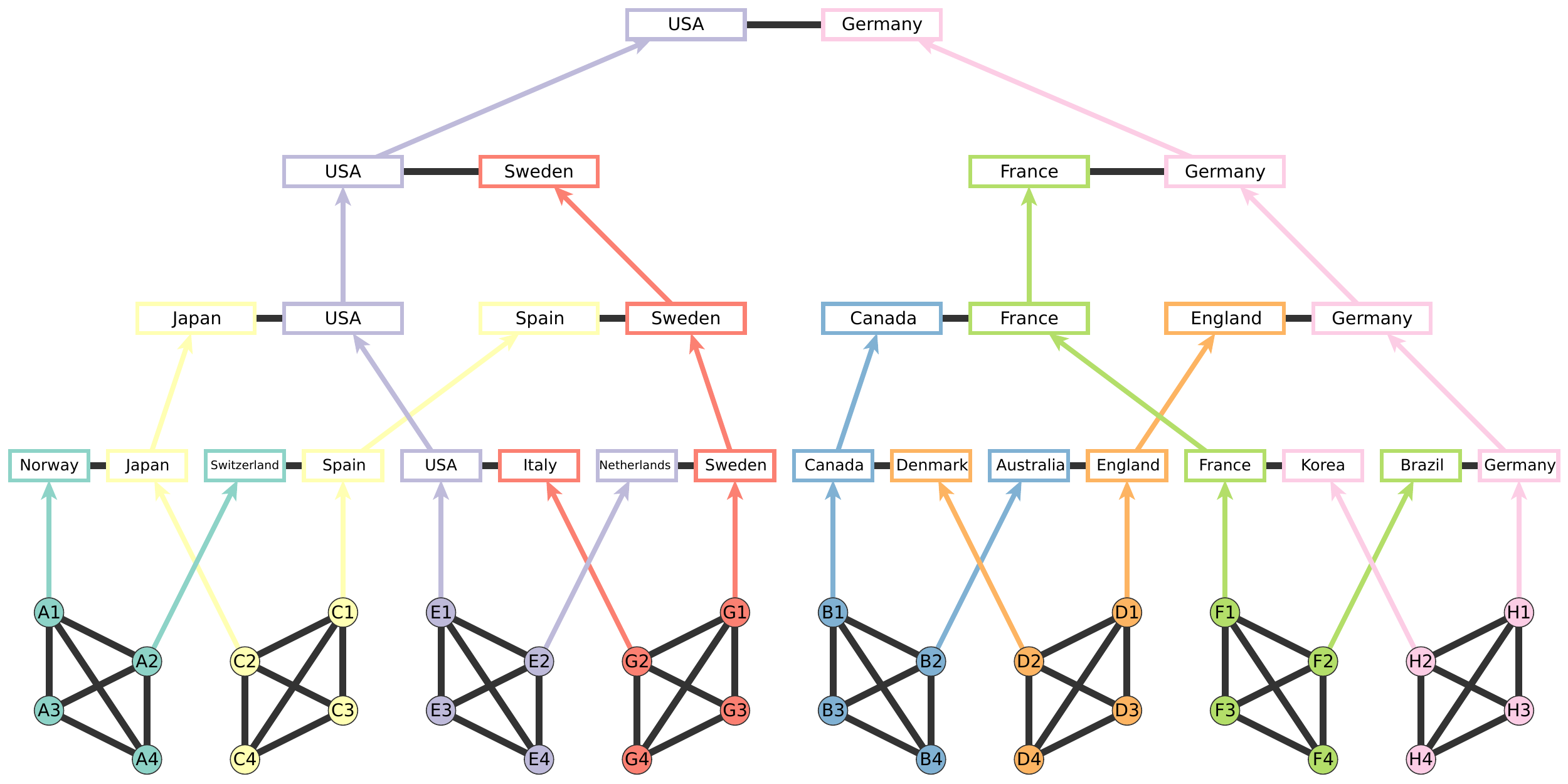}
\caption{Most likely bracket for FIFA Women's World Cup 2023
as predicted by current FIFA ranking.}
\label{fig:wwc2023-mostlikelytree}
\end{figure}

In Figure~\ref{fig:wwc2023-mostlikelytree},
the most likley tournament bracket is shown.
Like the winning probabilities,
it is a consequence of our use of the most recent FIFA ranking prior to the tournament
and should not be taken too seriously.

\begin{figure}[tb]
\begin{subfigure}{0.5\linewidth}
  \includegraphics[width=\linewidth]{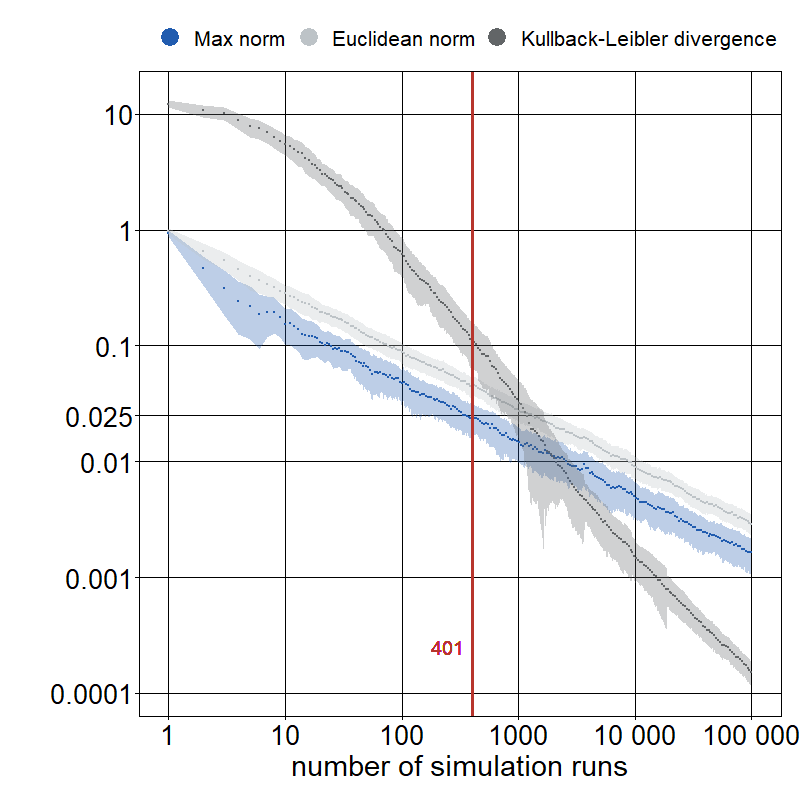}
    \caption{FIFA Men's World Cup 2022}
\end{subfigure}
\begin{subfigure}{0.5\linewidth}
  \includegraphics[width=\linewidth]{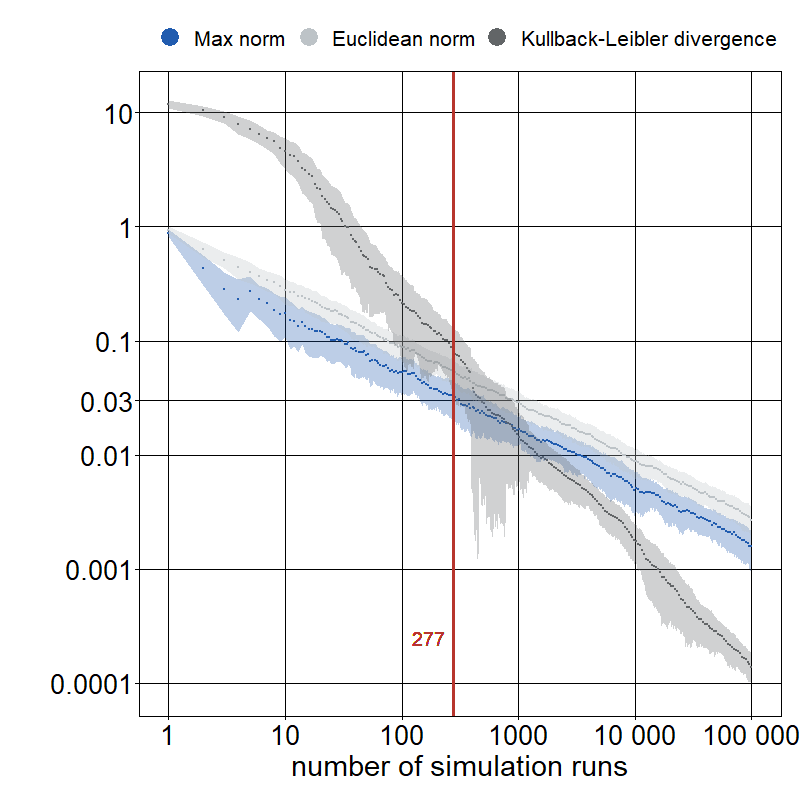}
    \caption{FIFA Women's World Cup 2023}
\end{subfigure}
\caption{Approximation error in probability estimates
obtained from relative frequencies in simulated tournaments (log-log scale).
For 100~independent trials, the median and middle quartiles are shown
for three different error measures with respect to true probabilities.
The red vertical line indicates the median number of simulation runs completed 
in the time needed for exact computation.
}
\label{fig:approximation}
\end{figure}

Since our winning probabilities largely reflect the FIFA ranking
and to some degree the expected difficulty of a team's schedule,
the above only served to demonstrate that results are plausible, given the assumptions. 
The real interest is in the relative computational efficiency 
of our exact algorithm with respect to simulation-based approximation.

We therefore ran tournament simulations with the same outcome model
and compare the quality of approximation to the (now available) exact probabilties
in Figure~\ref{fig:approximation}.
When the exact algorithm finishes,
the maximum error in a probability estimate is around $2.5\%$ points,
and thus larger than the majority of entries. 
To reduce it to even $1\%$ point, at least 10,000 simulations are necessary,
no less than 20,000 simulations are required for reasonable accuracy,
and even after 100,000 simulations the maximum error is still above $0.1\%$ points.

\section{Conclusion}\label{sec:conclusion}

We have shown how to exploit limited mixing in sports tournament brackets
to compute winning probabilities exactly.
The computation is independent of the model used for single-match outcomes,
and for the current FIFA World Cup schedules it is two orders of magnitude faster
than any reasonably accurate approximation through tournament simulation.

This should make our approach interesting for the assessment of
tournament-level consequences of differences in single-match prediction models,
and hence also for experiments with different parameter settings.
Note also that the algorithm can be run on partial tournaments
by setting the probabilities of known outcomes to~$1$.
Updates after any number of matches played,
or projections from current scores can thus be computed live.

The speed-up is facilitated by
limited mixing of team trajectories in the tournament design.
In each elimination round,
the number of groups which opposing teams may originate from
is no more than doubling.
Our approach therefore generalizes to tournament settings
in which integer parameters $t>t_1,t_2>0$
describe the number~$t$ of elimination rounds
after $2^{t_1}$ teams advance
and $2^{t_2}$ exit at the end of the group stage
($t=4$ and $t_1=t_2=1$ for the FIFA World Cup Finals above):
\begin{itemize}
\item $n=2^t\cdot(1+2^{t_2-t_1})$ teams
\item $m=2^{t-t_1}$ equal-sized groups,
	and therefore $\frac{n}{m}=2^{t_1}+2^{t_2}$ teams per group
\item $2^{t_1}$ teams advance from each of the $m$~groups,
	so that $\log(m\cdot 2^{t_1})=t$~elimination rounds are needed to determine the winner
\item groups $N_0,\ldots,N_{m-1}$ are ordered such that
	an $r$th-round elimination match between $i\in N_g$ and $j\in N_h$
	is possible only if $g\div2^{t_1+r-1}=h\div2^{t_1+r-1}$
\end{itemize}
The last item constrains the degree of mixing.

Running times are still prohibitive, however, if group sizes~$\frac{n}{m}$ are large. 
The FIFA Men's World Cup 2026 is planned for $n=48$~participants.
If they were to be organized in $m=8$~groups of $n/m=6$~teams each,
with $2$~teams from each group advancing into the elimination round,
then $t_1=1$ and $t_2=2$, and therefore $t=4$.
The number of matches would increase to
$m\cdot\binom{6}{2}=240$ in the group stage
and $2^t=16$ elimination matches (including a match for third place).
Each team plays five matches during the group stage
and up to four in the elimination rounds.
Not only does this add up to an enormous tournament of $256$~matches,
but there are $3^{\binom{6}{2}}=3^{15}=14,348,907$ possible outcome sequences to enumerate
in the group stage alone.

If participants are instead organized into $12$~groups of $4$~teams each,
the format does not fit our generalization,
because the number of groups is not a power of~$2$.
It is unlikely that a second stage with $8$~groups of $3$~teams 
will be used to set up a round of the last~16.
The more common format consists of $8$~of the $12$~third-placed teams advancing
to complete an elimination round of the last~32.
This would yield a total of $12\cdot\binom{4}{2}=72$ matches in the group stage 
and $2^t=32$ elimination matches (including a match for third place).
Each team plays three matches during the group stage
and up to five in the elimination rounds.
While a tournament with $104$~matches is almost twice as big as the current format,
the greater challenge for the computation of winning probabilities will lie
in the higher degree of trajectory mixing caused by third-placed teams
from a subset of groups not known in advance.

For the current format (which may soon be used for the UEFA EURO Finals),
and even for tournaments with $n=40$ teams in $8$~groups
or $n=64$ teams in $16$~groups,
both the number of matches and running time of exact computation
scale reasonably well with~$n$.

\bibliographystyle{apalike}
\bibliography{main.bib}

\begin{thebibliography}{}

\bibitem[Appleton, 1995]{appleton_may_1995}
Appleton, D.~R. (1995).
\newblock May the {Best} {Man} {Win}?
\newblock {\em The Statistician}, 44(4):529.

\bibitem[Batarfi and Reade, 2021]{batarfi_why_2021}
Batarfi, M. and Reade, J. (2021).
\newblock Why are {We} {So} {Good} {At} {Football}, and {They} {So} {Bad}?
  {Institutions} and {National} {Footballing} {Performance}.
\newblock {\em De Economist}, 169(1):63--80.

\bibitem[Bernholt et~al., 1999]{bernholt_football_1999}
Bernholt, T., Gülich, A., Hofmeister, T., and Schmitt, N. (1999).
\newblock Football {Elimination} {Is} {Hard} to {Decide} {Under} the
  3-{Point}-{Rule}.
\newblock In Kutyłowski, M., Pacholski, L., and Wierzbicki, T., editors, {\em
  Mathematical {Foundations} of {Computer} {Science} 1999}, Lecture {Notes} in
  {Computer} {Science}, pages 410--418, Berlin, Heidelberg. Springer.

\bibitem[Bernholt et~al., 2002]{bernholt_komplexitatstheorie_2002}
Bernholt, T., Gülich, A., Hofmeister, T., Schmitt, N., and Wegener, I. (2002).
\newblock Komplexitätstheorie, effiziente {Algorithmen} und die {Bundesliga}.
\newblock {\em Informatik-Spektrum}, 25(6):488--502.

\bibitem[Bettisworth et~al., 2023]{bettisworth_phylourny_2023}
Bettisworth, B., Jordan, A.~I., and Stamatakis, A. (2023).
\newblock Phylourny: efficiently calculating elimination tournament win
  probabilities via phylogenetic methods.
\newblock {\em Statistics and Computing}, 33(4):80.

\bibitem[Bradley and Terry, 1952]{bradley_rank_1952}
Bradley, R.~A. and Terry, M.~E. (1952).
\newblock Rank {Analysis} of {Incomplete} {Block} {Designs}: {I}. {The}
  {Method} of {Paired} {Comparisons}.
\newblock {\em Biometrika}, 39(3/4):324--345.
\newblock Publisher: [Oxford University Press, Biometrika Trust].

\bibitem[Cea et~al., 2020]{cea_analytics_2020}
Cea, S., Durán, G., Guajardo, M., Sauré, D., Siebert, J., and Zamorano, G.
  (2020).
\newblock An analytics approach to the {FIFA} ranking procedure and the {World}
  {Cup} final draw.
\newblock {\em Annals of Operations Research}, 286(1-2):119--146.

\bibitem[Chung and Hwang, 1978]{chung_stronger_1978}
Chung, F. R.~K. and Hwang, F.~K. (1978).
\newblock Do {Stronger} {Players} {Win} {More} {Knockout} {Tournaments}?
\newblock {\em Journal of the American Statistical Association},
  73(363):593--596.

\bibitem[David, 1959]{david_tournaments_1959}
David, H.~A. (1959).
\newblock Tournaments and {Paired} {Comparisons}.
\newblock {\em Biometrika}, 46(1/2):139.

\bibitem[David, 1988]{david_method_1988}
David, H.~A. (1988).
\newblock {\em The method of paired comparisons}.
\newblock Number no. 41 in Monograph. C. Griffin ; Oxford University Press,
  London : New York, 2nd ed., rev edition.

\bibitem[Davidson and Beaver, 1977]{davidson_extending_1977}
Davidson, R.~R. and Beaver, R.~J. (1977).
\newblock On {Extending} the {Bradley}-{Terry} {Model} to {Incorporate}
  {Within}-{Pair} {Order} {Effects}.
\newblock {\em Biometrics}, 33(4):693--702.

\bibitem[Edwards, 1991]{edwards_combinatorial_1991}
Edwards, C.~T. (1991).
\newblock {\em The combinatorial theory of single-elimination tournaments}.
\newblock PhD thesis, Montana State University.

\bibitem[Edwards, 1996]{edwards_double-elimination_1996}
Edwards, C.~T. (1996).
\newblock Double-{Elimination} {Tournaments}: {Counting} and {Calculating}.
\newblock {\em The American Statistician}, 50(1):27--33.

\bibitem[\'El{\H o}, 2008]{elo_rating_2008}
\'El{\H o}, A.~E. (2008).
\newblock {\em The {Rating} of {Chess} {Players}, {Past} and {Present}}.
\newblock Ishi Press, Bronx, NY.

\bibitem[Engist et~al., 2021]{engist_effect_2021}
Engist, O., Merkus, E., and Schafmeister, F. (2021).
\newblock The {Effect} of {Seeding} on {Tournament} {Outcomes}: {Evidence}
  {From} a {Regression}-{Discontinuity} {Design}.
\newblock {\em Journal of Sports Economics}, 22(1):115--136.
\newblock Publisher: SAGE Publications.

\bibitem[Glenn, 1960]{glenn_comparison_1960}
Glenn, W.~A. (1960).
\newblock A {Comparison} of the {Effectiveness} of {Tournaments}.
\newblock {\em Biometrika}, 47(3/4):253.

\bibitem[Groll et~al., 2015]{groll_prediction_2015}
Groll, A., Schauberger, G., and Tutz, G. (2015).
\newblock Prediction of major international soccer tournaments based on
  team-specific regularized {Poisson} regression: {An} application to the
  {FIFA} {World} {Cup} 2014.
\newblock {\em Journal of Quantitative Analysis in Sports}, 11(2):97--115.
\newblock Publisher: De Gruyter.

\bibitem[Hartigan, 1966]{hartigan_probabilistic_1966}
Hartigan, J.~A. (1966).
\newblock Probabilistic {Completion} of a {Knockout} {Tournament}.
\newblock {\em The Annals of Mathematical Statistics}, 37(2):495--503.

\bibitem[Horvat, 2020]{horvat_paul_2020}
Horvat, J. (2020).
\newblock From \emph{Paul} the {Octopus} to \emph{Achilles} the {Cat} –
  {Proper} {Names} of {Animals} which {Predict} the {Outcomes} of {Sports}
  {Competitions}.
\newblock {\em Folia Onomastica Croatica}, 29:73--121.

\bibitem[Horvat and Job, 2020]{horvat_use_2020}
Horvat, T. and Job, J. (2020).
\newblock The use of machine learning in sport outcome prediction: {A} review.
\newblock {\em WIREs Data Mining and Knowledge Discovery}, 10(5):e1380.

\bibitem[Hubáček et~al., 2022]{hubacek_forty_2022}
Hubáček, O., Šourek, G., and železný, F. (2022).
\newblock Forty years of score-based soccer match outcome prediction: an
  experimental review.
\newblock {\em IMA Journal of Management Mathematics}, 33(1):1--18.

\bibitem[Hwang, 1977]{hwang_several_1977}
Hwang, F.~K. (1977).
\newblock Several problems on knockout tournaments.
\newblock In Hoffman, F., Lesniak-Foster, L., and McCarthy, D., editors, {\em
  Proceedings of the {Eighth} {Southeastern} {Conference} on {Combinatorics},
  {Graph} {Theory}, and {Computing}: {Baton} {Rouge}, {Louisiana}, {February}
  28 - {March} 3, 1977}, volume~19 of {\em Congressus {Numerantium}}, pages
  363--380. Utilitas Mathematica Publishing.

\bibitem[Kern and Paulusma, 2001]{kern_new_2001}
Kern, W. and Paulusma, D. (2001).
\newblock The new {FIFA} rules are hard: complexity aspects of sports
  competitions.
\newblock {\em Discrete Applied Mathematics}, 108(3):317--323.

\bibitem[Koning et~al., 2003]{koning_simulation_2003}
Koning, R.~H., Koolhaas, M., Renes, G., and Ridder, G. (2003).
\newblock A simulation model for football championships.
\newblock {\em European Journal of Operational Research}, 148(2):268--276.

\bibitem[Kuper and Szymanski, 2022]{kuper_soccernomics_2022}
Kuper, S. and Szymanski, S. (2022).
\newblock {\em Soccernomics (2022 {World} {Cup} {Edition}): {Why} {France} and
  {Germany} {Win}, {Why} {England} {Is} {Starting} to and {Why} {The} {Rest} of
  the {World} {Loses}}.
\newblock HarperCollins, 2022nd edition edition.

\bibitem[Lepschy et~al., 2020]{lepschy_success_2020}
Lepschy, H., Wäsche, H., and Woll, A. (2020).
\newblock Success factors in football: an analysis of the {German}
  {Bundesliga}.
\newblock {\em International Journal of Performance Analysis in Sport},
  20(2):150--164.

\bibitem[Marchand, 2002]{marchand_comparison_2002}
Marchand, {\'E}. (2002).
\newblock On the comparison between standard and random knockout tournaments.
\newblock {\em Journal of the Royal Statistical Society: Series D (The
  Statistician)}, 51(2):169--178.

\bibitem[Maurer, 1975]{maurer_most_1975}
Maurer, W. (1975).
\newblock On {Most} {Effective} {Tournament} {Plans} {With} {Fewer} {Games}
  than {Competitors}.
\newblock {\em The Annals of Statistics}, 3(3).

\bibitem[Mcgarry and Schutz, 1997]{mcgarry_efficacy_1997}
Mcgarry, T. and Schutz, R.~W. (1997).
\newblock Efficacy of traditional sport tournament structures.
\newblock {\em Journal of the Operational Research Society}, 48(1):65--74.

\bibitem[Narayana and Zidek, 1969]{narayana_contributions_1969}
Narayana, T.~V. and Zidek, J. (1969).
\newblock Contributions to the {Theory} of {Tournaments} {Part~I:} {The}
  {Combinatorics} of {Knock}-{Out} {Tournaments}.
\newblock {\em Cahiers du Bureau universitaire de recherche opérationnelle
  Série Recherche}, 13:3--18.
\newblock Publisher: Institut Henri Poincaré - Institut de Statistique de
  l'Université de Paris.

\bibitem[Scarf et~al., 2009]{scarf_numerical_2009}
Scarf, P., Yusof, M.~M., and Bilbao, M. (2009).
\newblock A numerical study of designs for sporting contests.
\newblock {\em European Journal of Operational Research}, 198(1):190--198.

\bibitem[Scarf and Yusof, 2011]{scarf_numerical_2011}
Scarf, P.~A. and Yusof, M.~M. (2011).
\newblock A numerical study of tournament structure and seeding policy for the
  soccer {World} {Cup} {Finals}: {Tournament} design for the soccer {World}
  {Cup} {Finals}.
\newblock {\em Statistica Neerlandica}, 65(1):43--57.

\bibitem[Schauberger and Groll, 2018]{schauberger_predicting_2018}
Schauberger, G. and Groll, A. (2018).
\newblock Predicting matches in international football tournaments with random
  forests.
\newblock {\em Statistical Modelling}, 18(5-6):460--482.

\bibitem[Schwenk, 2000]{schwenk_what_2000}
Schwenk, A.~J. (2000).
\newblock What {Is} the {Correct} {Way} to {Seed} a {Knockout} {Tournament}?
\newblock {\em The American Mathematical Monthly}, 107(2):140--150.

\bibitem[Schwertman et~al., 1991]{schwertman_probability_1991}
Schwertman, N.~C., McCready, T.~A., and Howard, L. (1991).
\newblock Probability {Models} for the {NCAA} {Regional} {Basketball}
  {Tournaments}.
\newblock {\em The American Statistician}, 45(1):35--38.

\bibitem[Searls, 1963]{searls_probability_1963}
Searls, D.~T. (1963).
\newblock On the {Probability} of {Winning} with {Different} {Tournament}
  {Procedures}.
\newblock {\em Journal of the American Statistical Association},
  58(304):1064--1081.

\bibitem[Szczecinski and Roatis, 2022]{szczecinski_fifa_2022}
Szczecinski, L. and Roatis, I.-I. (2022).
\newblock {FIFA} ranking: {Evaluation} and path forward.
\newblock {\em Journal of Sports Analytics}, 8(4):231--250.
\newblock Publisher: IOS Press.

\bibitem[Sziklai et~al., 2022]{sziklai_efficacy_2022}
Sziklai, B.~R., Biró, P., and Csató, L. (2022).
\newblock The efficacy of tournament designs.
\newblock {\em Computers \& Operations Research}, 144:105821.

\bibitem[Tsokos et~al., 2019]{tsokos_modeling_2019}
Tsokos, A., Narayanan, S., Kosmidis, I., Baio, G., Cucuringu, M., Whitaker, G.,
  and Király, F. (2019).
\newblock Modeling outcomes of soccer matches.
\newblock {\em Machine Learning}, 108(1):77--95.

\bibitem[Winston et~al., 2022]{winston_mathletics_2022}
Winston, W.~L., Nestler, S., and Pelechrinis, K. (2022).
\newblock {\em Mathletics: {How} {Gamblers}, {Managers}, and {Fans} {Use}
  {Mathematics} in {Sports}, {Second} {Edition}}.
\newblock Princeton University Press, Princeton, 2nd edition edition.

\end{thebibliography}

\clearpage\appendix
\section{Team Strengths in FIFA Rankings}

Points in FIFA/Coca-Cola Rankings prior to tournament (rounded to nearest integer).

\bigskip\noindent
\begin{minipage}[t]{0.4\linewidth}
\begin{tabular}[t]{lr}
\multicolumn{2}{l}{\bf Men's (October 2022)}\\\hline
Brazil & 1841 \\
Belgium & 1817 \\
Argentina & 1774 \\
France & 1760 \\
England & 1728 \\
Spain & 1715\\
Netherlands & 1695\\
Portugal & 1677\\
Denmark & 1667\\
Germany & 1650\\
Croatia & 1646\\
Mexico & 1645\\
Uruguay & 1639\\
Switzerland & 1636\\
USA & 1627\\
Senegal & 1584\\
Wales & 1570\\
IR Iran & 1565\\
Serbia & 1564\\
Morocco & 1564\\
Japan & 1560\\
Poland & 1549\\
Ecuador & 1464\\
Korea Republic & 1530\\
Tunisia & 1508\\
Costa Rica & 1504\\
Australia & 1489\\
Canada & 1475\\
Cameroon & 1471\\
Qatar & 1440\\
Saudi Arabia & 1438\\
Ghana & 1393\\
\end{tabular}
\end{minipage}
\begin{minipage}[t]{0.4\linewidth}
\begin{tabular}[t]{lr}
\multicolumn{2}{l}{\bf Women's (June 2023)}\\\hline
USA & 2090 \\
Germany & 2062\\
Sweden & 2050\\
England & 2041\\
France & 2027\\
Spain & 2002\\
Canada & 1996\\
Brazil & 1995\\
Netherlands & 1980\\
Australia & 1920\\
Japan & 1917\\
Norway & 1908\\
Denmark & 1866\\
China PR & 1854\\
Italy & 1847\\
Korea Republic & 1840\\
Switzerland & 1766\\
Portugal & 1745\\
Republic of Ireland & 1744\\
Colombia & 1703\\
New Zealand & 1700\\
Argentina & 1682\\
Vietnam & 1649\\
Costa Rica & 1597\\
Nigeria & 1555\\
Jamaica & 1537\\
Philippines & 1513\\
Panama & 1483\\
Haiti & 1475\\
South Africa & 1472\\
Morocco & 1334\\
Zambia & 1298
\end{tabular}
\end{minipage}

\section{World Cup Groups}

\subsection{FIFA Men's World Cup 2022}\label{sec:mwcgroups}
\begin{tabular}{l|l|l|l}
\textbf{A} ($N_0$) &
\textbf{B} ($N_1$) &
\textbf{C} ($N_2$) &
\textbf{D} ($N_3$) \\\hline
Qatar & England & Argentina & France \\ 
Ecuador & IR Iran & Saudi Arabia & Australia \\ 
Senegal & USA & Mexico & Denmark \\ 
Netherlands & Wales & Poland & Tunisia \\ 
\multicolumn{4}{c}{~}\\
\textbf{E} ($N_4$) &
\textbf{F} ($N_5$) &
\textbf{G} ($N_6$) &
\textbf{H} ($N_7$) \\\hline
Spain & Belgium & Brazil & Portugal\\
Costa Rica & Canada & Serbia & Ghana\\
Germany & Morocco & Switzerland & Uruguay\\
Japan & Croatia & Cameroon & Korea Republic
\end{tabular}

\subsection{FIFA Women's World Cup 2023}\label{sec:wwcgroups}

\begin{tabular}{l|l|l|l}
\textbf{A} ($N_0$) &
\textbf{C} ($N_1$) &
\textbf{E} ($N_2$) &
\textbf{G} ($N_3$) \\\hline
New Zealand & Spain & USA & Sweden \\
Norway & Costa Rica & Vietnam & South Africa \\
Philippines & Zambia & Netherlands & Italy \\
Switzerland & Japan & Portugal & Argentina \\
\multicolumn{4}{c}{~}\\
\textbf{B} ($N_4$) &
\textbf{D} ($N_5$) &
\textbf{F} ($N_6$) &
\textbf{H} ($N_7$) \\\hline
Australia & England & France & Germany \\ 
Republic of Ireland & Haiti & Jamaica & Morocco \\ 
Nigeria & Denmark & Brazil & Colombia \\ 
Canada & China PR & Panama & Korea Republic
\end{tabular}

\section{Group Ranking Table}
$R[0,\ldots,728]$ to be added as supplementary information.

\section{Code}
Implementation in R to be provided for use with any single-match model.

\end{document}